\documentclass[preprint,review,3p]{elsarticle}
\usepackage{xcolor}
\usepackage{epsfig,epstopdf,color,caption,subcaption}
\usepackage{graphics,graphicx,amssymb,amsmath,verbatim}
\usepackage{amsbsy,mathrsfs,amsfonts,natbib,siunitx}
\usepackage[colorlinks=true, urlcolor=blue, citecolor=blue, linkcolor=blue]{hyperref}
\usepackage{multirow,enumerate}
\usepackage[process=all]{pstool}
\newcommand*{\QEDA}{\null\nobreak\hfill\ensuremath{\blacksquare}}
\newcommand*{\R}{\mathbb{R}}%
\usepackage{mathtools}
\graphicspath{{./Figures/}}
\DeclareMathOperator {\He}{He}

\DeclareMathOperator {\col}{col}

\DeclareMathOperator {\rk}{rank}
\newtheorem{mylemma}{Lemma}
\newtheorem{myproperty}{Property}
\newdefinition{myremark}{Remark}
\newproof{proof}{Proof}
\newtheorem{myassumptions}{Assumption}
\newtheorem{mytheorem}{Theorem}

\newtheorem{problem}{Problem}
\renewcommand{\Re}{\mathbb{R}}
\renewcommand{\epsilon}{\in}
\def\T {\top}
\def\expc {\mathbb{E}}
\allowdisplaybreaks
\journal{European Journal of Control. DOI: \href{https://doi.org/10.1016/j.ejcon.2023.100866}{https://doi.org/10.1016/j.ejcon.2023.100866}}
\begin{document}
\title{Output Regulation of Stochastic Sampled-Data Systems with Post-processing Internal Model}
\author[1]{Himadri Basu}\corref{cor1}%
\ead{himadri.basu@gipsa-lab.fr}

\author[2]{Francesco Ferrante}
\ead{francesco.ferrante@unipg.it}

\author[1]{Mirko Fiacchini\fnref{fn2}}
\ead{mirko.fiacchini@gipsa-lab.fr}

\cortext[cor1]{Corresponding author}
\fntext[fn2]{Research is funded in part by ANR via project HANDY, number ANR-18-CE40-0010.}

\address[1]{Univ. Grenoble Alpes, CNRS, Grenoble INP, GIPSA-Lab, Grenoble 38000, France}

\address[2]{Department of Engineering, University of Perugia, Perugia, Italy 06123}

\begin{abstract}
This paper deals with the output regulation problem (ORP) of a linear time-invariant (LTI) system in the presence of sporadically sampled measurement streams with the inter-sampling intervals following a stochastic process. Under such sporadically available measurement streams, a regulator consisting of a hybrid observer, continuous-time post-processing internal model, and stabilizer are proposed, which resets with the arrival of new measurements. The resulting system exhibits a deterministic behavior except for the jumps that occur at random sampling times and therefore the overall closed-loop system can be categorized as a piecewise deterministic Markov process (PDMP). In existing works on ORPs with aperiodic sampling, the requirement of boundedness on inter-sampling intervals precludes extending the solution to the random sampling intervals with possibly unbounded support. Using the Lyapunov-like theorem for the stability analysis of stochastic systems, we offer sufficient conditions to ensure that the overall closed-loop system is mean exponentially stable (MES) and the objectives of the ORP are achieved under stochastic sampling of measurement streams. The resulting LMI conditions lead to a numerically tractable design of the hybrid regulator. Finally, with the help of an illustrative example, the effectiveness of the theoretical results are verified.
\end{abstract}
\begin{center}
\textbf{This is an archival version of our paper. \\Please cite the published version DOI: \href{https://doi.org/10.1016/j.ejcon.2023.100866}{https://doi.org/10.1016/j.ejcon.2023.100866}}
\end{center}
\maketitle

\section{Introduction}
\subsection{Background}
The objective of the output regulation problem (ORP), also known as the servomechanism problem, is to control a specific output of the plant to track a prescribed reference trajectory and reject undesired disturbances, both of which being generated by an exosystem, while keeping all the trajectories of the system bounded \cite{14,Huang5}. Traditionally, there are two basic approaches for solving an ORP, namely the (i) feed-forward design approach which relies on the solution to regulator equations \cite{14} and (ii) internal model control approach \cite{8} that converts an output ORP to a stabilization problem that can be solved with simple eigenvalue placement method. In contrast with \cite{14,8} where the measured output of the plant is continuous, we consider the case where measurement streams are only available sporadically. In such cases, an emulation-based approach was adopted for both sampled linear systems \cite{Astolfi5,Huang5} and nonlinear systems \cite{Astolfi6}. 

In the presence of measurement intermittency for networked control systems, the emulation-based regulator ensures the closed-loop system trajectories and regulated output are bounded when the sampling interval is smaller than a certain threshold \cite{Astolfi5, Huang5}. With the design of a generalized hold device and internal model controller, a less conservative bound on the regulated output was derived for the same problem in \cite{Wang4}. Depending on the topology to connect the internal model and stabilizing units for the ORP with aperiodic sampling, in general, there are two types of architectures used for the ORP- namely (i) pre-processing internal model \cite{Wang4, Wang9} and (ii) post-processing internal model \cite{Astolfi6, Bin14,astolfi2013nonlinear}. 

In the first case, the internal model acts as a pre-processor on the control input and is driven by the stabilizing control action, while in the second case, the internal model directly processes the regulated error signals and generates an input for the stabilizer \cite{Bin14}. For single-input-single-error (SISE) systems, both schemes are fundamentally equivalent with pre-processing internal model architecture being more constructive in some cases such as in \cite{Wang4,Carolis11,Wang9}, while post-processing internal model simplifies the regulator design \cite{Basu2022,Basu2021}. Using either or both of these internal model paradigms, the ORP was studied by the authors in \cite{Carolis11,Lawrence9,Fujioka10} for periodically sampled measurement updates and in \cite{Basu2022,Wang4,Huang5} for uncertain, time-varying and aperiodic sampling intervals.

In all of the papers, mentioned above, the solution to the ORP with measurement intermittency requires the inter-sampling intervals to be bounded from above. However, this assumption may not be realistic. Some recent works have addressed the stabilization of linear sampled-data systems subject to a random sampling interval \cite{tanwani:hal-01962130,Huff} with possibly unbounded support. Inspired by the works \cite{Huff,antunes2}, we extend our previous results on ORP under bounded aperiodic sampling \cite{Basu2022} to unbounded random sampling. Since the solution to the ORP in \cite{Wang4,Wang9,Huang5,Astolfi5} requires the sampling intervals to be bounded, the results cannot be immediately adopted for the case of random time-varying intervals. Although the ORP has been studied for stochastic linear systems, for example in \cite{He7,Scarciotti}, it was still assumed that the measurement streams are available continuously. 

As noted above, the ORP has been extensively studied for continuous stochastic systems and aperiodic sampled-data systems with bounded inter-sampling intervals, see for example \cite{Huang5} and the references therein, but to the best of the authors' knowledge there are not many analogous results for the same problem with stochastically sampled measurement streams.

\subsection{Contributions} The main contribution of this work is to derive a regulator to ensure that the plant-stabilizer closed-loop system is internally stable and the regulated output is mean asymptotically convergent to zero in the presence of stochastically available measurement streams. The regulator comprises a continuous-time stabilizer, post-processing internal model, and hybrid observer which resets with the arrival of new measurement streams and reconstructs the regulated output from such stochastically available data. As in \cite{tanwani:hal-01962130,Huff}, we consider that the intervals between consecutive measurement samples undergo a Poisson process, \emph{i.e.} the sampling intervals form a sequence of independent and identically distributed random variables with exponential distribution. In contrast with the works of \cite{Astolfi5,Huang5,Huff} for networked control systems where the measurement output between the samples is held constant before being processed, the hybrid observer proposed in this work implicitly acts as a generalized hold device for the reconstruction of the regulated output. 

As the resulting closed-loop system exhibits deterministic flow dynamics interrupted by the jumps occurring at random sampling times, it is classified as a PDMP. Using Dykin's equation \cite[pg.~31]{Davis} and the stochastic version of Lyapunov-like theorem, the sufficient stability conditions guaranteeing mean exponentially stable (MES) of the closed-loop system yield a set of linear matrix inequalities (LMI) which consequently lead to the straightforward computation of regulator parameters. With the help of a numerical case study, we highlight the effectiveness of the proposed approach. 

The paper is organized as follows. Section \ref{sec:2} contains the basic definitions and introduces the problem formulation. In Section \ref{sec:3}, we review the post-processing internal model architecture for stochastically available measurement streams. In Section \ref{sec:4}, we present the main results related to sufficient conditions for MES of the closed-loop system and the mean asymptotic convergence of the regulated output. A numerical example to illustrate the effectiveness of the proposed approach is presented in Section \ref{sec:5}. Finally, some concluding remarks are given in Section \ref{sec:6}.

\subsection{Notations}
$P[x]$ denotes probability and $\mathbb{E}[x]$ expectation of a random variable $x$. The set $\mathbb{N}$ denotes the set of positive integers including zero. $I$ and ${0}$ represent respectively an identity matrix and a zero matrix with appropriate dimensions. $\mathbb{R}^{n\times{m}}$ denotes a space of real matrices with order $n\times{m}$. An open right-half complex plane is denoted by $\mathbb{C}^{+}$. For a symmetric matrix $A$, $A>0\ (\text{or}\ A\geq{0})$ denotes that matrix $A$ is positive definite (or semi-definite). For square matrices $A_i,i=1,2,\cdots,N$ of compatible dimensions, $A=\text{diag}(A_1,A_2,\cdots,A_N)$ represents a block-diagonal matrix with diagonal elements $A_i$. The notation $\lambda_{\text{max}}(A)\ (\text{or}\ \lambda_{\text{min}}(A))$ represents the maximal (or minimal) real part of the eigenvalues of matrix $A$, $\text{He}(A)=A+A^\text{T}$, $\sigma(A)$ denotes the spectrum of eigenvalues of $A$. For $x,y\in\mathbb{R}^{N}$, $\|x\|$ is the Euclidean norm measuring the distance of $x$ from the origin, $\text{col}(x,y)=\begin{bmatrix}x^{\text{T}},y^{\text{T}}\end{bmatrix}^{\T}$. A shorthand notation $x^{+}$ is used to denote the jump $x(t^{+})$ at sampling time instants. 
\section{Problem Formulation}\label{sec:2}
Consider a linear time-invariant plant of the following form
\begin{equation}
\begin{aligned}\label{eq:plant}
\mathcal{P}\begin{cases}
\dot{x}_{p} &= A_{p}x_p+B_{p}u+E_{p}w,\\
y_{p} &= C_{p}x_p,\\
e_{p} &= y_{p}- y_{w},\\
y_w &= F_{p}w,
\end{cases}
\end{aligned}
\end{equation}
with $x_{p}\in\Re^{n_p},~u\in\Re^{m_p},~y_p,~y_w,~e_p\in\Re^{p}$ being respectively the state, the control law to be designed, the measured output of the plant, exosystem output to be tracked, and the error signal which is thus required to be regulated to zero. The exogenous signal $w\in\Re^{q}$ is generated by an exosystem of the form 
\begin{equation}\label{eq:2}
\dot{w}=Sw,
\end{equation}
where the exosystem matrix $S$ is assumed to be neutrally stable, \emph{i.e.} $S$ has all eigenvalues on the imaginary axis. While $S$ is perfectly known, the exosystem state $w$ in \eqref{eq:2} is not directly available for feedback design. The matrices $A_p, B_p, E_p, C_p$ and $F_p$ in \eqref{eq:plant} are constant matrices of appropriate dimensions. The output $y_p$ is available only at some isolated time instances $t_k,\ k\in{\mathbb{N}}$ with inter-sampling intervals $\delta_k=t_{k+1}-t_k,\ k=1,2,\cdots,\infty$. In this work, we assume that $\{\delta_k\}$ is a sequence of independent and identically distributed random variables with exponential distribution
\begin{equation}\label{eq:exp}
F(s)=P[\delta_k\leq{s}]=1-e^{-\lambda{s}},k\in\mathbb{N},s\geq{0},
\end{equation}
where $\lambda>0,\ \expc[\delta_k]=\dfrac{1}{\lambda}$. If the number of sampling events occurred until the current time $t$ is denoted by
\begin{equation}\label{eq:poisson}
N_t=\sup\{k\in\mathbb{N}|t_k\leq{t}\},    
\end{equation}
then the probability of $N_t=n$ under the Poisson process of intensity $\lambda$ is given by
\begin{equation}\label{eq:poisson2}
P[N_t=n]=e^{-\lambda{t}}\dfrac{(\lambda{t})^n}{n!}.    \end{equation}
Inter-sampling events of a stochastic sampled-data system have been successfully modeled with a Poisson process in \cite{Huff,Tabbara}. For $n=\infty$, $P[N_t=\infty]=0$ from \eqref{eq:poisson2}, \emph{i.e.} there is zero probability of an infinite number of sampling events occurring up until finite time $t$. Furthermore, with $s=0$, $F(s)=0$ in \eqref{eq:exp} implying zero probability of two sampling events occurring concurrently with zero dwell-time. These two facts discard the possibility of Zeno behavior. 

Let us now introduce the following assumptions for the solvability of ORP \cite{14}. 

\begin{myassumptions}\label{as:1}
The matrix pair $(A_p,B_p)$ is stabilizable and $(A_p,C_p)$ is detectable.\hfill $\diamond$
\end{myassumptions}

\begin{myassumptions}\label{as:2}
The matrix $\begin{bmatrix}A_p-\lambda{I} & B_p\\C_p & 0\end{bmatrix},\ \lambda\in\sigma(S)$ is of full rank, or equivalently there exists a unique solution pair $(X_p,R)$ to the following linear regulator equations
\begin{equation}\label{eq:4}
X_pS=A_pX_p+B_p{R}+E_p,\ C_pX_p-F_p=0,
\end{equation} 
where the matrix $X_p$ uniquely defines the time-invariant manifold $X_p{w}$ of the plant state $x_p$, on which the regulated output $e_{p}=0$. Additionally, the steady-state input $u=Rw$ renders the given manifold $\mathbb{E}[x_p]=X_pw$ positively invariant. \hfill $\diamond$
\end{myassumptions}

As noted in \cite{14}, under Assumptions \ref{as:1}, 
 and \ref{as:2}, the classical ORP (continuous availability of measurement) for the plant \eqref{eq:plant} is solvable by a dynamic error feedback control of the form 
\begin{equation}
\begin{aligned}\label{eq:5}
u &={K}z,~
\dot{z}={G}_{1}z+{G}_{2}e_p,
\end{aligned}
\end{equation}
where $z\in\Re^{n_z}$ is the regulator state to be specified later and the constant controller gain matrices ${G}_1$ and ${G}_{2}$ with appropriate dimensions are defined as in \cite{Basu2022}. In the presence of sporadic availability of measurement streams, $e_p$ in \eqref{eq:5} should be replaced with a reconstructed form of $e_p$, \emph{i.e.} $\hat{e}_p$ which is generated by a hybrid observer discussed in the next section. The internal model state vector $z$ approaches its steady state $Zw$ for the solvability of the ORP, where $Z\in\mathbb{R}^{q\times{n_z}}$ satisfies
\begin{equation}\label{eq:int_model}
ZS=G_1{Z}, KZ=R,    
\end{equation}
by virtue of the internal model principle \cite{14}. With the traditional output regulation framework being laid out, we are now ready to define the objectives of the stochastic ORP considered in this paper. 
\begin{problem}\label{prob:1}
(i) When $w=0$, the origin of the unperturbed closed-loop system is mean exponentially stable (MES), \emph{i.e.}
there exist two positive scalars $c, \gamma_{0}$ such that for every initial condition ${x}_{p}(0)={x}_{p0}\in\mathbb{R}^{n_p}$,
\begin{equation}\label{eq:MES1}
\mathbb{E}[\|x_p(t)\|^2]\leq{c}e^{-\gamma_{0}t}\|x_{p0}\|^2
\end{equation}
with $\gamma_{0}>0$ being the decay rate of the trajectories; and (ii) the closed-loop system remain internally stable when $w\neq{0}$ with mean regulated output asymptotically converging to zero, \emph{i.e.} $\lim_{t\to\infty}\mathbb{E}[e(t)]=0$.
\end{problem}

\section{ Solution Outline of Post-processing Internal Model}\label{sec:3}
\begin{figure}
\centering
\includegraphics[width=0.65\textwidth]{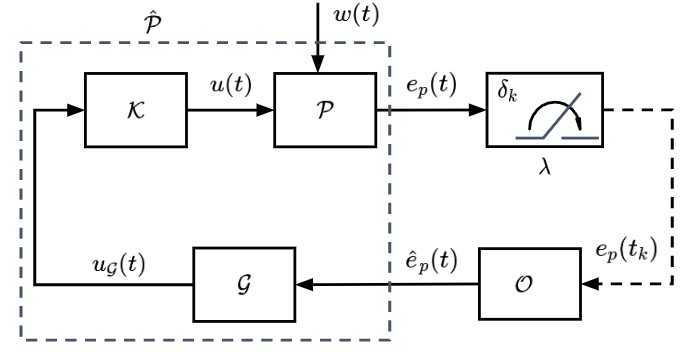}
\caption{{\label{fig.1a}} Schematic block diagram of the ``post-processing'' internal model architecture for LTI plant with sporadic measurements (average sampling rate is $\lambda$). Continuous-time signals are marked with solid arrows, while the sporadic measurements are with dashed arrows.}\vspace{-2ex}
\end{figure}

Since the output of the plant is sporadically available, we propose a control scheme depicted in Figure \ref{fig.1a}, which is constituted by a post-processing internal model $\mathcal{G}$, hybrid observer $\mathcal{O}$, and a stabilizing controller $\mathcal{K}$. The hybrid estimator $\mathcal{O}$ is designed to provide a converging estimate of the regulation error $\hat{e}_p$ from the intermittent measurements as follows
\begin{equation}\label{eq:h.o}
\mathcal{O}\begin{cases}
\begin{aligned}
&\dot{\chi} = \mathsf{T}\chi,\ \hspace{14ex} \text{if}~t\neq{t_k},\\
&\chi^{+} =\mathsf{L_1}\chi(t)+\mathsf{L_2}e_p(t),\ \text{if}~t={t_k},\\
&\hat{e}_{p} = \mathsf{H}\chi,
\end{aligned}\end{cases}
\end{equation}
where $\chi\in\R^{n_{\chi}}$, observer matrices $\mathsf{L}_i,~i=1,2$, and $\mathsf{T}$ to be designed later. This estimated regulated output $\hat{e}_{p}$ is then fed as an input to the internal model
\begin{equation}\label{eq:int_model2}
\mathcal{G}\begin{cases}
\begin{aligned}
&\dot{z}=G_{1}z+G_{2}\hat{e}_{p},\ \ \text{if}~t\neq{t_k},\\
&z^{+}= z(t),\ \hspace{7ex}\text{if}~t={t_k},\\
&u_{\mathcal{G}}= Kz,
\end{aligned}\end{cases}
\end{equation} 
where $z\in\mathbb{R}^{n_z}$. The output of the internal model $u_{\mathcal{G}}$ then acts as an input to a continuous-time stabilizer $\mathcal{K}$ with state vector $\zeta\in\mathbb{R}^{n_{\zeta}}$ of the form
\begin{equation}\label{eq:stab}
\mathcal{K}\begin{cases}\begin{aligned}
&\dot{\zeta} = A_{\zeta}\zeta+B_{\zeta}{u_{\mathcal{G}}},\ \text{if}~t\neq{t_k},\\
&\zeta^{+} = \zeta,\ \hspace{8.3ex}\ \text{if}~t={t_k},\\
& u = C_{\zeta}\zeta+D_{\zeta}u_{\mathcal{G}}.
\end{aligned}\end{cases}
\end{equation}
As we shall show later, the design of such stabilizer parameters can be computed by simple eigenvalue-placement techniques aimed at stabilizing the cascaded closed-loop system $\hat{\mathcal{P}}$. 
Before moving forward, we let the following property hold.

\begin{myproperty}\label{assu:prop1}
Let 
\begin{align*}
A_{\text{cl}} &= \begin{bmatrix}A_p & B_pC_{\zeta}\\0 & A_{\zeta}\end{bmatrix},B_{\text{cl}}=\begin{bmatrix}B_{p}D_{\zeta}\\B_{\zeta}\end{bmatrix}K, \mathsf{H}_{1}=\begin{bmatrix}C_p & 0\end{bmatrix},\nonumber
\end{align*}\vspace{-1ex}
then there exist a solution $(X_M, Z)$ to the equations 
\begin{equation}\label{eq:Francis}
X_MS=A_{\text{cl}}X_M+B_{\text{cl}}{Z}+E_{\text{cl}},\ \mathsf{H}_{1}X_M-F_p=0,
\end{equation} 
for all $E_{\text{cl}}\in\mathbb{R}^{(n_p+n_{\zeta})\times{q}}$ and $F_p\in\mathbb{R}^{p\times{q}}$,  where $E_{\text{cl}}= \begin{bmatrix}E^{\T}_p\!& \!0\end{bmatrix}^{\T}, \text{and}\ Z$ satisfies $ZS=G_{1}Z$ in \eqref{eq:int_model}.$\hfill \diamond$
\end{myproperty}

We show later that \text{Property}~\ref{assu:prop1} is fulfilled under some natural conditions. Let us define an augmented stabilizer-plant system with the state vector $x_{M}=\text{col}(x_p,\zeta)$ as   
\begin{equation}
\begin{aligned}
& \dot{x}_{M}= A_{cl}x_{M}+B_{cl}z+E_{cl}w,\ \text{if}~t\neq{t_k},\\
& e_{p}= \mathsf{H}_{1}x_{M}-F_{p}w,\\
& x_{M}^{+}= x_{M},\hspace{18.5ex}\text{if}~t={t_k},
\end{aligned}
\end{equation}
where internal model state $z$ evolves according to \eqref{eq:int_model2} and is rewritten as
\begin{equation}
\begin{aligned}
&\dot{z}=G_{1}z+G_{2}\mathsf{H}_{1}\chi,\ \text{if}~t\neq{t_k},\\
&z^{+}=z,\hspace{11.5ex}\text{if}~t={t_k}.
\end{aligned}
\end{equation}
By using the change of coordinates $\tilde{x}_{M}=x_{M}-X_{M}w, \tilde{z}=z-Zw$ and defining $\tilde{x}_{\alpha}=\text{col}(\tilde{x}_{M},\tilde{z})$ with $Z,X_{M}$ being solutions to \eqref{eq:int_model} and \eqref{eq:Francis} respectively, the augmented system $\hat{\mathcal{P}}$ consisting of the stabilizer-plant and internal model dynamics in transformed coordinates yields the following form
\begin{equation}\label{eq:eq40}
\begin{aligned}
& \dot{\tilde{x}}_{\alpha} = \mathfrak{A}_{c}\tilde{x}_{\alpha}+\mathfrak{B}_{c}(\mathsf{H}\chi-\begin{bmatrix}\mathsf{H}_{1}&0\end{bmatrix}\tilde{x}_{\alpha}),\hspace{2.4ex}\text{if}~t\neq{t_k},\\
& \tilde{x}_{\alpha}^{+} = \tilde{x}_{\alpha},\hspace{28.2ex}\text{if}~t={t_k}, 
\end{aligned}
\end{equation}
where 
\begin{equation}
\begin{aligned}
\mathfrak{A}_{c}= \mathfrak{A}\!+\!\mathfrak{B}_{c}\begin{bmatrix}\mathsf{H}_{1}&0\end{bmatrix}, \mathfrak{A}=\begin{bmatrix}A_{\text{cl}} & B_{\text{cl}}\\0 & G_{1}\end{bmatrix}, \mathfrak{B}_{c} = \begin{bmatrix}0\\G_{2}\end{bmatrix}.\nonumber
\end{aligned}
\end{equation}
At this stage, it is clear that to solve the ORP, $\mathsf{T}$ and $\mathsf{L}_{i},~i=1,2$ in \eqref{eq:h.o} need to be designed to ensure that $\hat{e}_p\to{e}_p$ or in other words $\tilde{e}_p\coloneqq (e_p-\hat{e}_p)$ approaches zero asymptotically. To this end, we denote 
$\chi\in\mathbb{R}^{n_{\chi}}=(\chi_1, \chi_2)$, where $n_{\chi}=n+p\ \text{and}\ n=n_p+n_{\zeta}+n_z$. The state component $\chi_{1}$ can be viewed as an estimate of $\tilde{x}_{\alpha}$ in \eqref{eq:eq40} and $\chi_{2}$ of $\tilde{e}_{p}$. Based on this, let us now select the observer parameters as
\begin{equation}\label{eq:eq45}
\begin{aligned}
\mathsf{T} & = \begin{bmatrix}\mathfrak{A}_{c} & Q\\0 & W\end{bmatrix}, \mathsf{L}_{1}=\begin{bmatrix}I & 0\\-\mathsf{H}_{2} & 0\end{bmatrix}, \mathsf{L}_{2}=\begin{bmatrix}0\\I\end{bmatrix},\\
\mathsf{H}_{2} &=\begin{bmatrix}\mathsf{H}_{1} & 0\end{bmatrix}, \mathsf{H}=\begin{bmatrix}\mathsf{H}_{2}&0\end{bmatrix},
\end{aligned}
\end{equation}
where $Q\in\R^{n\times p}$ and $W\in\R^{p\times p}$ are to be designed. 
\begin{myproperty}\label{prop:2}
Let the state matrix $\mathfrak{A}_c$ of the extended plant model $\hat{\mathcal{P}}$ in transformed coordinates \eqref{eq:eq40} be Hurwitz. Then, the Property \ref{assu:prop1} holds.   
\end{myproperty}
\begin{proof}
it suffices to notice that if 
$$\mathfrak{A}_c=\mathfrak{A}+\mathfrak{B}_{c}\mathsf{H}_{2}$$
is Hurwitz, then the triple
$\left(\mathfrak{A},\mathfrak{B}_{c},\mathsf{H}_{2}\right)$
is stabilizable and detectable. Therefore, one has that for all $\lambda_{\mu}\in\mathbb{C}_+$, the matrix 
$$
\begin{aligned}
\rk&\begin{bmatrix}
A_{\text{cl}}-\lambda_{\mu} I&B_{\text{cl}}&0\\
0&G_1-\lambda_{\mu} I&G_2
\end{bmatrix}=n\\
\rk&\begin{bmatrix}
A_{\text{cl}}-\lambda_{\mu} I&B_{\text{cl}}\\
0&G_1-\lambda_{\mu} I\\
\mathsf{H}_{1}&0
\end{bmatrix}=n+p
\end{aligned}
$$
From the second condition above, it turns out that for all $\lambda_{S}\in\sigma(G_1)=\sigma(S)\subset\mathbb{C}_+$
$$
\rk\begin{bmatrix}
A_{\text{cl}}-\lambda_{S} I&B_{\text{cl}}\\
\mathsf{H}_{1}&0\\
0&G_1-\lambda_{S} I
\end{bmatrix}=n+p,
$$
which guarantees the existence of a solution pair $(X_{M},Z)$ to \eqref{eq:Francis} in Property~\ref{assu:prop1} by Theorem 1.9 of \cite{14}.  \QEDA
\end{proof}
 
We now illustrate how to select the stabilizer parameters $A_{\zeta},B_{\zeta},C_{\zeta},D_{\zeta}$ such that the matrix $\mathfrak{A}_{c}$ in \eqref{eq:eq40} becomes Hurwitz. Let us define a non-singular matrix $T_{\zeta}\in\mathbb{R}^{n\times{n}}$ that transforms $\mathfrak{A}_{c}$ as follows
\begin{equation}\label{eq:bar}
\bar{\mathfrak{A}}_{c}\ =\ T^{-1}_{\zeta}\mathfrak{A}_{c}T_{\zeta}\ =\ \begin{bmatrix}A_p & B_pD_{\zeta}K & B_{p}C_{\zeta}\\G_{2}C_{p} & G_{1} & 0\\0 & B_{\zeta}K & A_{\zeta}\end{bmatrix},\ T_{\zeta}=\begin{bmatrix}I_{n_p} & 0 & 0\\0 & 0 & I_{n_\zeta}\\0 & I_{n_z} & 0\end{bmatrix}. 
\end{equation}
With a little abuse of notation, let us denote $x=\col{(x_e,\hat{x}_{e})}\in\mathbb{R}^{n}$, and introduce a system dynamics of the form 
\begin{equation}\label{eq:19bar}
\dot{x}\ =\ \bar{\mathfrak{A}}_{c} {x},    
\end{equation}
where $\bar{\mathfrak{A}}_{c}$ is given in \eqref{eq:bar}. For $\mathfrak{A}_{c}$ to become Hurwitz, it is both necessary and sufficient that $\bar{\mathfrak{A}}_{c}$ in \eqref{eq:bar} is Hurwitz, or equivalently $x(t)\to\ 0$ as $t\to\infty$. To do this, let us first rewrite the dynamics \eqref{eq:19bar} as
\begin{align}
\dot{x}_{e} & = \underbrace{\begin{bmatrix}A_p & B_pD_{\zeta}K\\G_{2}C_p & G_1\end{bmatrix}}_{A}\ x_{e} + \underbrace{\begin{bmatrix}B_p\\0\end{bmatrix}}_{B}C_{\zeta} \hat{x}_{e},\label{eq:20bar}\\
\dot{\hat{x}}_{e} & = A_{\zeta} \hat{x}_{e}+B_{\zeta}\underbrace{\begin{bmatrix}0 & K\end{bmatrix}}_{C} x_{e},\label{eq:21bar}
\end{align}
where $\hat{x}_{e}$ is viewed as an estimate of $x_{e}$ with $B_{\zeta}$ as the observer gain, and $C_{\zeta}\hat{x}_{e}$ as the observer-based feedback control law to be designed.

A general observer structure to reconstruct the state $x_e$ that evolves according to \eqref{eq:20bar} is given as follows
\begin{equation}
\dot{\hat{x}}_{e}=(A+BC_{\zeta})\hat{x}_{e}+B_{\zeta}C\left(x_{e}-\hat{x}_{e}\right)=\underbrace{\left(A+BC_{\zeta}-B_{\zeta}C\right)}_{A_\zeta}\hat{x}_{e}+B_{\zeta}Cx_{e}    
\end{equation}
The overall system \eqref{eq:19bar} can be equivalently expressed as decoupled stabilization and estimation problems by virtue of the separation principle. Based on this, the design conditions are enumerated as follows:
\begin{itemize}
\item select controller gain $C_{\zeta}$ such that the matrix $A+BC_{\zeta}$ is Hurwitz,
\item select an observer gain $B_{\zeta}$ such that the matrix $A-B_{\zeta}C$ is Hurwitz, and
\item determine $A_{\zeta}=A+BC_{\zeta}-B_{\zeta}C$.
\end{itemize}
Furthermore, the convergence rate of (i) the decoupled control \& estimation problems and consequently that of (ii) $x\to{0}$ in \eqref{eq:19bar} can be directly tuned by a suitable selection of gains $C_{\zeta}$ and $B_{\zeta}$.

\begin{myremark}
The system matrix $\mathfrak{A}_{c}$ of the extended plant model $\hat{\mathcal{P}}$ is thus made Hurwitz by the suitable selection of the stabilizer parameters $A_{\zeta},B_{\zeta},C_{\zeta},D_{\zeta}$ through a simple eigenvalue placement method. This is in contrast with the pre-processing architecture \cite{Basu2022} where the stabilizer parameters are computed simultaneously with other regulator parameters by solving a convex optimization problem. However, as noted in \cite{Basu2021,Basu2022}, the convexification of sufficient stability conditions to obtain LMIs in pre-processing paradigm requires intricate algebraic manipulations. $\hfill\square$
\end{myremark}

By defining $\tilde{\chi}_{1}=\tilde{x}_{\alpha}-\chi_{1}$ and $\tilde{\chi}_{2}=\chi_{2}-\mathsf{H}_{2}\tilde{\chi}_{1}$, the overall closed-loop system with augmented state variables $\tilde{x}=\text{col}(\tilde{x}_{\alpha},\tilde{\chi})\in\mathbb{R}^{2n+p}$, $\tilde{\chi}=\text{col}(\tilde{\chi}_{1},\tilde{\chi}_{2})$ can be described by the following impulsive system \cite{antunes2}
\begin{equation}\label{eq:17}
\begin{aligned}
\begin{cases}
& \dot{\tilde{x}}\!=\!\left[\begin{array}{c|c}\mathfrak{A}_{c} & -\mathfrak{B}_{c}\mathsf{H}\\\hline0 & M\end{array}\right]\!\tilde{x},\quad\forall{t}\geq{0},\ t\neq{t_k},\\
& \tilde{x}^{+} = \text{diag}(I_{2n},0_{p})\tilde{x},\hspace{5ex}t=t_k,
\end{cases}
\end{aligned}
\end{equation}
where 
$$M =\begin{bmatrix}\mathfrak{A}-Q\mathsf{H}_{2} & -Q\\-\mathsf{H}_{2}\mathfrak{A}+R\mathsf{H}_{2} & R\end{bmatrix}, R=W+\mathsf{H}_{2}Q.$$
With this post-processing internal model architecture, the ORP, defined in \text{Problem} \ref{prob:1} reduces to the stabilization problem of a stochastic linear sampled-data system with mean sampling rate $\lambda>0$ under Poisson distribution. Clearly, the objectives in \text{Problem} \ref{prob:1} are met if the equilibrium point $\tilde{x}=0$ in \eqref{eq:17} is MES. Before we end this section, let us redefine the problem statement we focus on in this paper.
\begin{problem}\label{prob:2}
Given the intensity $\lambda>0$ of the Poisson sampling process, design the regulator parameters $Q$ and $W$ such that the resulting closed-loop system \eqref{eq:17} is MES.
\end{problem}
\section{Main Results}\label{sec:4}
In this section, we will derive the stability conditions to solve \text{Problem} \ref{prob:2} for the closed-loop system \eqref{eq:17} which is deterministic except for jumps occurring at random sampling times. According to \cite{teel2}, system \eqref{eq:17} is a PDMP. This fact allows us to adopt the following key result from \cite{antunes2,Huff}, which is viewed as a stochastic analog to Lyapunov's stability theorems for deterministic nonlinear systems. Since the dynamics of $\tilde{\chi}$ in \eqref{eq:17} is decoupled from $\tilde{x}_{\alpha}$, we first need to show the MES of the following PDMP.
\begin{equation}\label{eq:19a}
\begin{aligned}
\begin{cases}
\dot{\tilde{\chi}}&\!=\ M\tilde{\chi}=f(\tilde{\chi}),\ t\neq{t_k},\\
\tilde{\chi}^{+}&\!=\ N\tilde{\chi}=g(\tilde{\chi}),\hspace{1ex} t={t_k}, 
\end{cases}
\end{aligned}
\end{equation}
where $N=\text{diag}(I_{n},0_{p})$.
\begin{mytheorem}\label{thm:1}
If $V:\mathbb{R}^{n+p}\to\mathbb{R}$ is a continuous differentiable function such that    
\begin{equation}\label{eq:19}
\mathbb{E}\left[\sum_{k\leq{N}_{p}}\left|V(\tilde{\chi}^{+})-V(\tilde{\chi})\right|\right]<\infty, p\in\mathbb{N}.   
\end{equation}
Then, for $t\geq{0}$ and $\forall{\tilde{\chi}(0)}=\tilde{\chi}_{0}\in\mathbb{R}^{n+p}$,
\begin{align}
\mathbb{E}[V(\tilde{\chi}(t))] & =V(\tilde{\chi}_{0})+\mathbb{E}\left[\int_{0}^{t}\mathfrak{U}V(\tilde{\chi}(s)\ ds)\right],\label{eq:21}\\    
\mathfrak{U}V(\tilde{\chi}) & = \nabla{V^{\T}}(\tilde{\chi})f(\tilde{\chi})+\lambda\left(V(g(\tilde{\chi})-V(\tilde{\chi})\right),\label{eq:22}
\end{align}
where $\nabla{V}(\tilde{\chi})$ is the gradient of $V(\tilde{\chi})$.
\end{mytheorem}

Next, we derive the sufficient conditions for MES of \eqref{eq:19a} from Theorem \ref{thm:1}, which leads to a computationally tractable design of regulator parameters $Q$ and $W$.
\begin{mylemma}
Given the Poisson sampling rate $\lambda>0$, the trajectories of \eqref{eq:19a} are MES with the decay rate $\gamma$, if $Q=P^{-1}_{1}\bar{Q}$, $W=P^{-1}_{2}\bar{R}-\mathsf{H}_{2}Q$ where $P_1\in\mathbb{R}^{n}>0$, $P_2\in\mathbb{R}^{p}>0$, $\bar{Q}\in\mathbb{R}^{n\times{p}}$, $\bar{R}\in\mathbb{R}^{p\times{p}}$, $\gamma>0$ are solutions to 
\begin{equation}\label{eq:23}
\hspace{-1.5ex}\left[\!\begin{array}{c|c}\!\He{(P_{1}\mathfrak{A}\!-\!\bar{Q}\mathsf{H}_{2})}\!+\!\gamma{P_1}\!&\!\! -\mathfrak{A}^{\T}\mathsf{H}^{\T}_{2}P_2\!+\!\mathsf{H}^{\T}_{2}\bar{R}^{\T}\!-\!\bar{Q}\!\!\\\hline\noalign{\vspace{1ex}}
\bullet \!\!&\!\! \He{(\bar{R})}\!+\!(\gamma-\!\lambda){P_2}
\end{array}\!\right]\!\leq\!{0}.   \vspace{1ex}
\end{equation}
\end{mylemma}

\begin{proof}
From the PDMP of the trajectories $\tilde{\chi}$ in \eqref{eq:19a}, for any $t\geq{0}$ we obtain
\begin{align}
&\tilde{\chi}(t)=\!e^{M(t\!-\!t_{N_t})}Ne^{M(t_{N_t}-t_{N_{t}-1})}\!\cdots\!{N}e^{M(t_2-t_1)}Ne^{Mt_1}\tilde{\chi}_{0},\nonumber\\
&\|\tilde{\chi}(t)\|\leq e^{c_{1}t}\|\tilde{\chi}_{0}\|,\ c_{1}=\|M\|\geq{0},\ \|N\|=1. 
\end{align}
Therefore, for any $k\geq{1}$, $\|\tilde{\chi}(t^{+}_k)\|\leq\|\tilde{\chi}(t_k)\|\leq\ e^{c_{1}t_k}\|\tilde{\chi}_{0}\|$. Let us first consider a time-varying energy function $\mathbf{V}(\tilde{\chi},t)=e^{\gamma{t}}\tilde{\chi}^{\T}P\tilde{\chi}$ with $P=\text{diag}(P_1,P_2)\in\mathbb{R}^{n+p}>0$. Then, for any $\tilde{\chi}_{0}\in\mathbb{R}^{n+p}$, we have 
\begin{align}
 & \mathbb{E}\left[\sum_{k\leq{N}_{p}}\left|\mathbf{V}(\tilde{\chi}^{+},t^{+}_k)-\mathbf{V}(\tilde{\chi},t_k)\right|\right]\ =\ \mathbb{E}\left[\sum_{k\leq{N}_{p}}e^{\gamma{t_k}}\left|\tilde{\chi}^{\T}(t^{+}_k)P\tilde{\chi}(t^{+}_k)-\tilde{\chi}^{\T}(t_k)P\tilde{\chi}(t_k)\right|\right]\nonumber\\
\leq\ & 2\|P\|\mathbb{E}\left[\sum_{k\leq{N_p}}\!\!\!e^{\gamma{t_k}}\|\tilde{\chi}(t_k)\|^2\right]\leq{c_2}\mathbb{E}\left[\sum_{k\leq{N}_{p}}\!\!\!e^{(2c_{1}+\gamma)t_k}\right]\ \leq\ c_{2}N_pe^{(2c_{1}+\gamma)p}\ <\infty,
\end{align}
where $c_2=2\|P\|\|\tilde{\chi}_{0}\|^2$. Then, according to Theorem \ref{thm:1}, we move on to evaluate \eqref{eq:21}. By differentiating $\mathbf{V}(\tilde{\chi},t)$ along the trajectories of the flow-map in \eqref{eq:19a} and adding the scaled difference of energy due to the jumps as in \eqref{eq:22}, $\mathfrak{U}\mathbf{V}(\tilde{\chi},t)=\tilde{\chi}^{\T}\mathcal{M}\tilde{\chi}$ yields:
\begin{equation}
\mathcal{M}=\left[\begin{array}{c|c}\He{(P_{1}(\mathfrak{A}-Q\mathsf{H}_{2}))}+\gamma{P_1}&-\mathfrak{A}^{\T}\mathsf{H}^{\T}_{2}P_{2}+\mathsf{H}^{\T}_{2}R^{\T}P_{2}-P_{1}Q\\\hline\noalign{\vspace{0.5ex}}\bullet \!&\!\He{(P_{2}Q)}+(\gamma-\!\lambda){P_2}\end{array}\right],\nonumber
\end{equation}
which after the substitution of $P_{1}Q=\bar{Q}$ and $P_{2}R=\bar{R}$ reduces to $\mathcal{M}\leq{0}$ by \eqref{eq:23}. Therefore, $\mathfrak{U}\mathbf{V}(\tilde{\chi},t)\leq{0}$ and consequently from \eqref{eq:21}, we obtain
\begin{align}
&\mathbb{E}\left[\mathbf{V}(\tilde{\chi},t)\right]\leq \mathbf{V}(\tilde{\chi}_{0})\implies{e^{\gamma{t}}}\mathbb{E}\left[\tilde{\chi}^{\T}P\tilde{\chi}\right]\leq\tilde{\chi}^{\T}_{0}P\tilde{\chi}_{0},
\end{align}
or equivalently
\begin{equation}\label{eq:27a}
\mathbb{E}\left[\|\tilde{\chi}(t)\|^2\right]\leq\dfrac{\lambda_{\max}(P)}{\lambda_{\min}(P)}e^{-\gamma{t}}\|\tilde{\chi}_{0}\|^2, \forall \tilde{\chi}_{0}\in\mathbb{R}^{n+p}.
\end{equation}
This concludes the proof. \QEDA
\end{proof}

Since the matrix $\mathfrak{A}_{c}$ in \eqref{eq:17} is Hurwitz, by using the MES property of the trajectories $\tilde{\chi}$ in \eqref{eq:19}, the origin of the continuous system dynamics concerning the state variable $\tilde{x}_{\alpha}$ in \eqref{eq:17} can be shown to be MES as well. This fact is illustrated next. Let us rewrite the closed-loop system with state variable $\tilde{x}_{\alpha}$ from \eqref{eq:17} as
\begin{equation}\label{eq:27}
\dot{\tilde{x}}_{\alpha}=\mathfrak{A}_{c}\tilde{x}_{\alpha}-\mathfrak{B}_{c}\mathsf{H}\tilde{\chi},
\end{equation}
where $\lambda_{\max}(\mathfrak{A}_{c})<0$ and $\|e^{\mathfrak{A}_{c}t}\|\leq c_{2}e^{-\beta{t}}$, for $c_{2},\beta>0$ by \text{Property} \ref{prop:2}. Furthermore, to achieve a satisfactory convergence rate $\beta>0$, the stabilizer parameters in \eqref{eq:stab} can be appropriately chosen, as discussed above. By using these facts along with Young's inequality and Jensen's inequality for integrals \cite{Gu3}, from \eqref{eq:27} we then obtain for any $t\geq{0}$
\begin{align}
&\mathbb{E}\left[\|\tilde{x}_{\alpha}\|^2\right]\leq {2}\|e^{\mathfrak{A}_{c}t}\|^2\|\tilde{x}_{\alpha{0}}\|^2+ 2t\|\mathfrak{B}_{c}\mathsf{H}\|\times\mathbb{E}\left[\int_{0}^{t}\hspace{-1ex}\|e^{\mathfrak{A}_{c}(t-s)}\|^2\|\tilde{\chi}(s)\|^2~ds\right],\nonumber\\
& \leq\!c_{3}e^{-2\beta{t}}\|\tilde{x}_{\alpha}(0)\|^2\!+\!c_{4}t\!\!\int_{0}^{t}\hspace{-1ex}e^{-2\beta(t-s)}\mathbb{E}\left[\|\tilde{\chi}(s)\|^2\right]ds,\label{eq:28}
\end{align}
where $c_{3}=2c^{2}_{2}$, $c_{4}=c_{3}\|\mathfrak{B}_{c}\mathsf{H}\|$. Next, the MES of \eqref{eq:19} and the scalar identity $te^{-\gamma{t}}\leq\dfrac{2}{e}e^{-(\gamma{t}/{2})}$ yields 
\begin{equation}\label{eq:30}
\mathbb{E}\left[\|\tilde{x}_{\alpha}\|^2\right]\leq\ c_{3}e^{-2\beta{t}}\|\tilde{x}_{\alpha}(0)\|^2+{c_{5}e^{-\gamma{t}/2}}\|\tilde{\chi}(0)\|^2,
\end{equation}
where $c_5=\dfrac{2c_{4}\lambda_{\max}(P)}{\lambda_{\min}(P)(2\beta-\gamma)e}$. Using $\|\tilde{x}\|^2\leq\|\tilde{x}_{\alpha}\|^2+\|\tilde{\chi}\|^2$ and the results in \eqref{eq:27a}, \eqref{eq:30}, we finally obtain
\begin{equation}\label{eq:31a}
\mathbb{E}\left[\|\tilde{x}\|^2\right]\leq{c}e^{-\gamma_{0}t}\|\tilde{x}_{0}\|^2
\end{equation}
where $c=\max{\left(\!c_{3},c_{5}\!+\!\dfrac{\lambda_{\max}(P)}{\lambda_{\min}(P)}\!\right)}, \gamma_{0}=\min{\left(2\beta,\dfrac{\gamma}{2}\right)}$. Thus the origin of the closed-loop system is MES and \text{Problem} \ref{prob:2} is solved. As a consequence, 
\begin{align}
\left(\mathbb{E}\left[\|e(t)\|\right]\right)^2 & \leq\mathbb{E}[\|e\|^2]\leq\|\mathsf{H}_{2}\|\mathbb{E}\left[\|\tilde{x}_{\alpha}\|^2\right]\leq\|\mathsf{H}_{2}\|\mathbb{E}\left[\|\tilde{x}\|^2\right]\leq{c}\|\mathsf{H}_{2}\|e^{-\gamma_{0}t}\|\tilde{x}_{0}\|^2,
\end{align}
or equivalently $\mathbb{E}\left[\|e(t)\|\right]\leq\sqrt{c\|\mathsf{H}_{2}\|}e^{-\left(\gamma_{0}/2\right)t}\|\tilde{x}_{0}\|$ and hence $\lim_{t\to\infty}\mathbb{E}[e(t)]=0$ for all $\tilde{x}_{0}\in\mathbb{R}^{2n+p}$.
\begin{myremark}
The conditions in \eqref{eq:23} is not directly LMI because of the nonlinear terms $\gamma{P_1}$ and $\gamma{P_2}$. However, for a fixed value of $\gamma$, the conditions in \eqref{eq:23} are LMIs and the maximum value of the decay rate $\gamma$ for which \eqref{eq:23} holds is computed through a simple line search. Furthermore, given a desired choice of decay rate ($\gamma>0$), we can even compute the maximum Poisson sampling rate $\lambda$ for which the closed-loop system \eqref{eq:19a} is MES by solving the following optimization problem
$$\underset{P_1,P_2,\bar{Q},\bar{R},\gamma}{\hspace{5.5ex}\text{max}\ \lambda}\ \text{subject to}\ \eqref{eq:23}.$$
\end{myremark}
\section{Illustrative Examples}\label{sec:5}
To illustrate the effectiveness of the post-processing internal model architecture in bestowing mean asymptotic convergence of the regulation error, let us now consider two numerical examples.

\textit{Example $1$:} In this example, we consider a tracking problem where a mono-frequency harmonic oscillator's states are to be tracked by an unstable second-order system despite disturbances. This example mimics the position-tracking problem of a networked servomotor under sporadically available measurements. Let us take the plant \eqref{eq:plant} and exosystem state matrices \eqref{eq:2} of the form
\begin{equation}
\begin{aligned}
A_p & =\begin{bmatrix}-2 &1\\0 & 0.8\end{bmatrix},\ B_p=\begin{bmatrix}0\\1\end{bmatrix},\ E_p\!=\!\begin{bmatrix}1 &0\\0 & 0\end{bmatrix}, C_p = \begin{bmatrix}0.1 & 0\end{bmatrix},\\
~F_p & =\begin{bmatrix}0 & 2\end{bmatrix},\ S = \begin{bmatrix}0 & 1\\-1 & 0\end{bmatrix},~\sigma(S)\ =\ \pm\ 1i. \label{eq:ex1}
\end{aligned}
\end{equation}
In this example, the plant is the servomotor model with the states $x_p=\col{(x_{\text{pos}},x_{\text{vel}})}$, where $x_{\text{pos}}$ is the position and $x_{\text{vel}})$ is the velocity of the motor. The objective is to make the servomotor position $x_{\text{pos}}$ track $10F_{p}w$.

\begin{figure}
\centering
\includegraphics[width=0.8\textwidth]{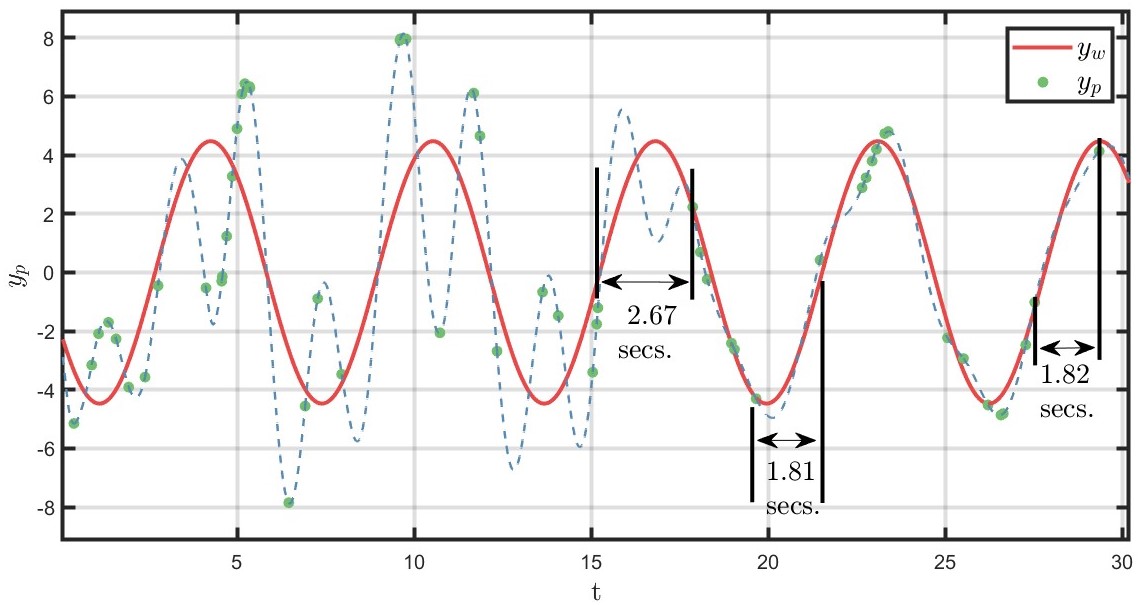}
\caption{{\label{fig.4}} $y_p$ tracks the exosystem measurement $y_w$. The sampled measurements available from the motor are marked in green while the non-available part of the measurements is marked with dashed lines}\vspace{-2ex}
\end{figure}

As shown in Figure~\ref{fig.4}, the output of the motor is available only intermittently and the inter-sampling intervals between consecutive measurements $y_p$ are randomly picked from an exponential distribution with an average sampling rate $\lambda=2$. The inter-sampling intervals, as evident from Figure~\ref{fig.4}, can be as high as $2.67$ seconds. 

\begin{figure}
  \centering
     \begin{subfigure}{0.53\textwidth}   
         \includegraphics[width=\textwidth]{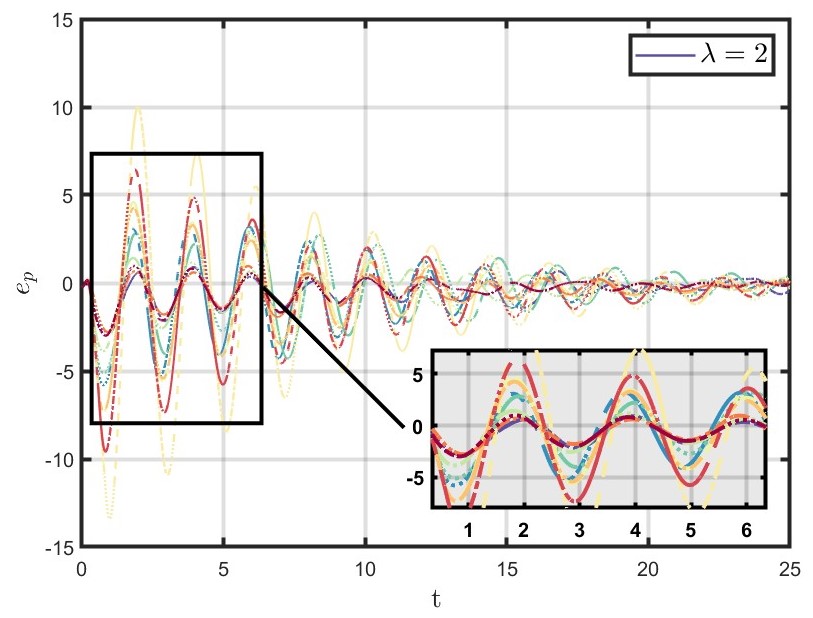}
         \caption{$e_p$ for average sampling rate $\lambda\ =\ 2$.}
         \label{fig.2a}
         \end{subfigure}
         \hfill
     \begin{subfigure}{0.45\textwidth}
         \centering
         \includegraphics[width=\textwidth]{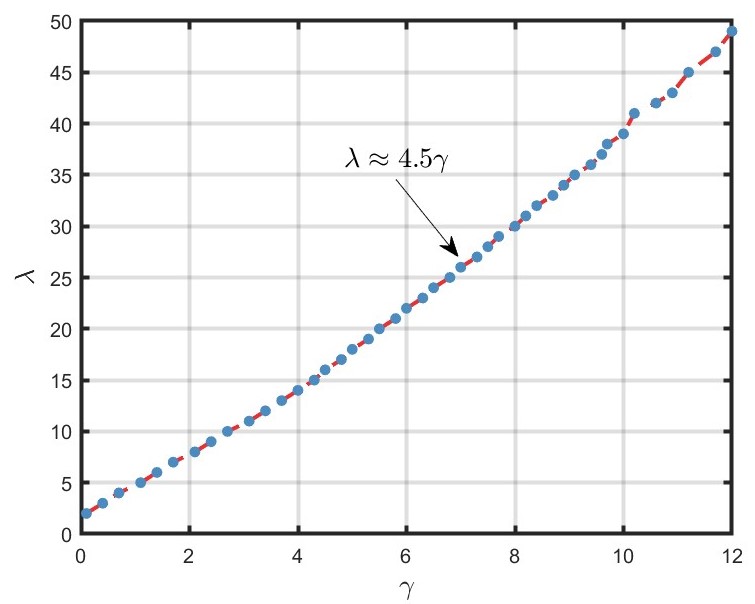}
         \caption{$\lambda(\gamma)$}
         \label{fig:lambda}
     \end{subfigure}
        \caption{Multiple numerical simulations showing the convergence of regulated output $e_p$ for $\lambda=2$, and the variation of the estimation decay rate $\gamma$ with $\lambda$ and vice -versa}
     \end{figure}

To achieve output regulation, we design a continuous-time stabilizer satisfying \eqref{eq:stab} with the stabilizer matrices given as follows
\begin{align}
A_{\zeta} & = \begin{bmatrix}-2 &\!\!1 &\!\!14.98 &\!\! 0 \\
 -5.85 &\!\!-2.8 &\!\!37.94 &\!\!16.5\\
-0.5 &\!0 &\!\!-3.6 &\!1 \\
-0.4 &\!0 &\! -2.24 &\!0 
\end{bmatrix},\ B_{\zeta} = \begin{bmatrix}-14.98 \\ -33.44 \\ 3.6 \\ 1.24 \end{bmatrix},\ C_{\zeta} = \begin{bmatrix}-5.85 \\ -3.6 \\ 4.5 \\ 16.5\end{bmatrix}^{\T},\ D_{\zeta}=0,\label{eq:57}
\end{align}
which makes the closed-loop system matrix $\mathfrak{A}_{c}$ in \eqref{eq:eq40} Hurwitz with decay rate $\beta=0.1$. 
It is easy to verify that the Assumptions \ref{as:1}, \ref{as:2} are satisfied. Then, according to \cite{14}, we select $1$- copy internal model of the exosystem \eqref{eq:int_model2} as
\begin{equation}\label{eq:eq36}
G_1=\begin{bmatrix}0 & 1\\-1 & 0\end{bmatrix}, G_2=\begin{bmatrix}-5\\-4\end{bmatrix}. 
\end{equation} 
From \eqref{eq:23}, the decay rate and the hybrid observer parameters are found to be
\begin{equation}\label{eq:QW}
\begin{aligned}
\gamma = 0.1, \ W & =-116.008,\ Q = \begin{bmatrix}Q_1 & Q_2 & Q_3\end{bmatrix}^{\T},\ Q_1 = \begin{bmatrix} 1163.51 & 3374.21 & 132.36\end{bmatrix},\\
Q_2 & = \begin{bmatrix}578.15 & 100.99 & 154.48\end{bmatrix},\ Q_3 =\begin{bmatrix}120.87 & 170.665\end{bmatrix}.
\end{aligned}
\end{equation}
Numerical solutions to LMIs are obtained using YALMIP toolbox \cite{Lofberg7} with SDPT3 solver \cite{tutuncu6} in Matlab\textsuperscript{\textcopyright}. With these regulator parameters, the closed-loop system matrix \eqref{eq:31a} is MES at the origin and the regulated output also asymptotically converges to zero as shown in Figure \ref{fig.2a} for different sampling sequences with average sampling rate $\lambda=2$. The decay rate of convergence $\gamma$ for $\hat{e}_{p}\to\ e_{p}$ by the hybrid observer $\mathcal{O}$ in \eqref{eq:h.o} is $0.1$. However, to obtain an even faster convergence (higher $\gamma$) and smaller settling time, a frequent measure of $y_p$ is required. This is evident from Figure \ref{fig:lambda} where a larger decay rate requirement on the performance measure can be attained by a proportional increment on the sampling rate $\lambda$. To yield a decay rate $\gamma\geq{0.1}$, the corresponding mean sampling rate to be selected is $\lambda\geq{2}$. Thus the objectives of the position synchronization problem of a networked servomotor are achieved under intermittent measurements with stochastic inter-sampling intervals. 

\begin{figure}
\centering
\includegraphics[width=0.8\textwidth]{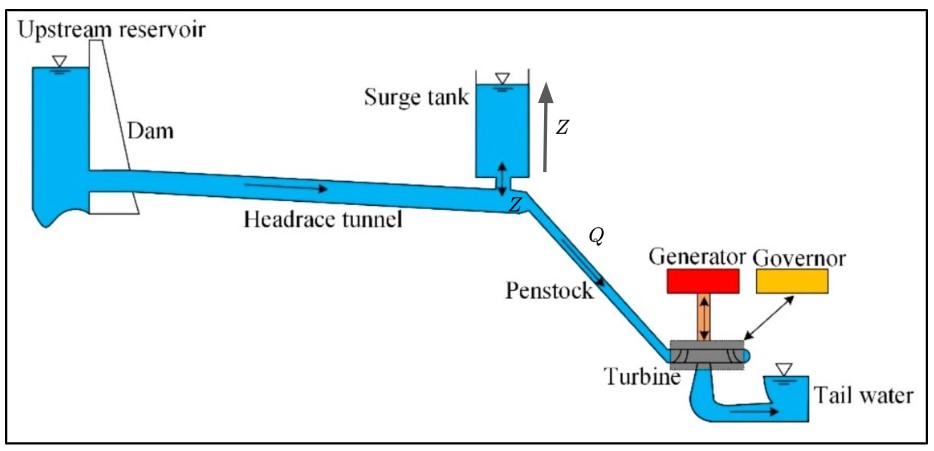}
\caption{{\label{fig.pen}} An abstract representation of hydroelectric power plant with reservoir, penstock, and turbine \cite{WinNT}}\vspace{-2ex}
\end{figure}

\begin{figure}
\centering
\includegraphics[width=0.8\textwidth]{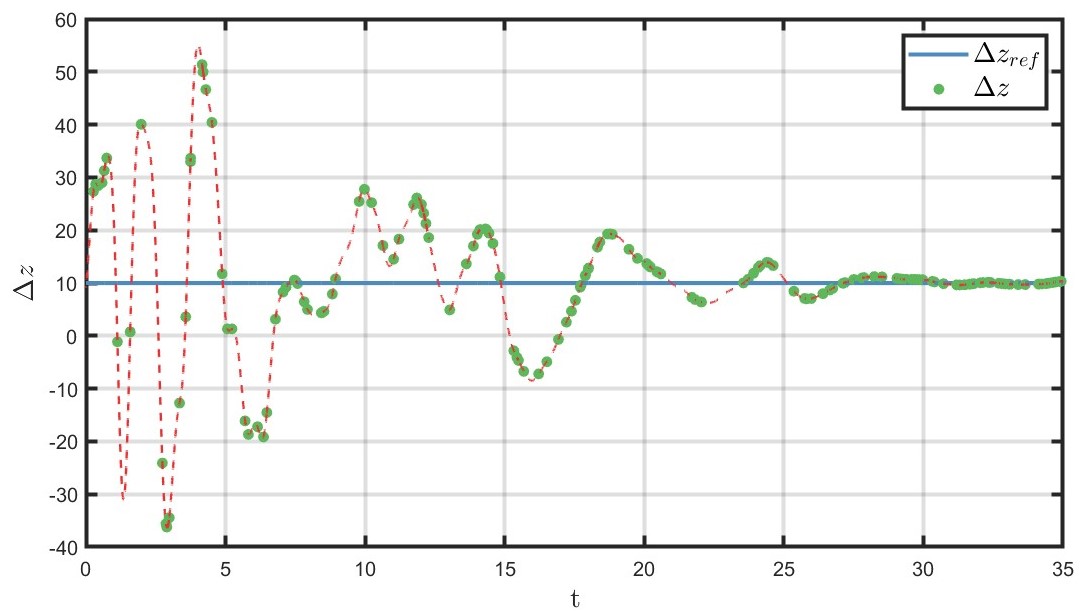}
\caption{{\label{fig.pen_stock}} Water level rise in surge tank $\Delta{z}\to\ \Delta{z}_{ref}$}\vspace{-2ex}
\end{figure}

\textit{Example 2:} Next, we consider an application from a hydroelectric power plant where a surge tank is required to maintain a desired water level by the appropriate control of flow through the penstock, as shown in Figure \ref{fig.pen}. The relationship between the water level rise in the surge tank $\Delta{z}$ and the flow rate in front of the turbine $\Delta{q}$ is given in \cite{Stuksrud4} by a second-order transfer function model 
\begin{equation}
\dfrac{\Delta{z}}{\Delta{q}}=\dfrac{k(s+\alpha_{1})}{s^{2}+\alpha_{2}s+\alpha_{3}}.
\end{equation}
By considering $k=0.5,\ \alpha_{1}=0.2,\ \alpha_{2}=-0.8,\ \alpha_{3}=-0.5$, and assuming a constant reference signal to be tracked, the plant model \eqref{eq:plant} and exosystem parameters \eqref{eq:2} are as follows
\begin{equation}
\begin{aligned}
A_p & =\begin{bmatrix}0 &1\\0.5 & 0.8\end{bmatrix},\ B_p=\begin{bmatrix}0\\1\end{bmatrix},\ E_p\!=\!\begin{bmatrix}1 &0\\0 & 0\end{bmatrix}, C_p = \begin{bmatrix}0.1 & 0.5\end{bmatrix},\\
~F_p & =\begin{bmatrix}0 & 2\end{bmatrix},\ S = \begin{bmatrix}0 & 1\\0 & 0\end{bmatrix},~\sigma(S)\ =\ \{0,0\}. \label{eq:ex2}
\end{aligned}
\end{equation}
The output of the plant, which in this case is the water-level rise, is available sporadically with a mean sampling rate $\lambda=4.5$. Despite the sporadic measurements, to achieve tracking in this case with a convergence rate $\gamma=0.1$, the regulator parameters are determined as follows
\begin{equation}
\begin{aligned}
Q & =\begin{bmatrix}67.41 & 104.11 & 191.66 & 73.36 & -4.05 & 55.95 & -41.84 & 0.3543\end{bmatrix}^{\T},\\
W & = -2998.8,\ G_{1} = S,\ G_{2} = \begin{bmatrix}2 & -4\end{bmatrix}^{\T},\\
A_{\zeta} &= \begin{bmatrix}0 &\!\!1 &\!\!25.42 &\!\! 0 \\
 -37 &\!\!-5.8 &\!\!32.31 &\!\!-8.87\\
0.2 &\!1 &\!\!-6.6 &\!1 \\
-0.4 &\!-2 &\! -52.43 &\!0 
\end{bmatrix},\ B_{\zeta} = \begin{bmatrix}-25.42 \\ -24.81 \\ 6.6 \\ 52.43\end{bmatrix},\ C_{\zeta} = \begin{bmatrix}-37.5 \\ -6.6 \\ 2.5 \\ -8.87\end{bmatrix}^{\T},\ D_{\zeta}=5.
\end{aligned}
\end{equation}
With these regulator parameters, the overall closed-loop system is MES at the origin and the regulated output asymptotically converges to zero. This can be seen from Figure \ref{fig.pen_stock} where the water-level rise in the surge tank eventually reaches the desired height with the prescribed rate of convergence. Therefore, our proposed approach to designing the regulator successfully renders MES of the overall closed-loop system even under sporadically sampled measurement streams.
\section{Conclusions}\label{sec:6}
In this paper, we have extended our previous results on aperiodically sampled ORP with bounded inter-sampling intervals to the case of stochastically sampled measurement streams with possibly unbounded support. With the use of a post-processing architecture consisting of a hybrid observer, internal model and a stabilizer, we achieve MES of the closed-loop system and mean asymptotic convergence of the regulated error. The internal model converts the original problem into a hybrid stabilization problem. Using a stochastic analog of Lyapunov's stability theorem, we offer sufficient LMI conditions for MES, which then leads to a numerically tractable regulator design.

Unlike our previous results, the proposed regulator is no longer dependent on the bounds of inter-sampling intervals. But the knowledge of average sampling rate is sufficient for this ORP problem to be solved under a Poisson sampling process. Building upon this result, our ongoing work focuses on solving cooperative ORP of multi-agent systems subject to sporadic sampling of transmitted information between the agents.

\bibliographystyle{IEEEtran}
\bibliography{ref}

\end{document}